%% file: paper.tex
\documentclass[12pt]{article}
\usepackage{epsfig,a4}
\def\be{\begin{equation}}
\def\ee{\end{equation}}
\def\bea{\begin{eqnarray}}
\def\eea{\end{eqnarray}}
\def\d{\mbox{d}}
\def\nn{\nonumber}

\begin{document}
\begin{titlepage}
\begin{flushright}
Cavendish-HEP-00/01\\
DAMTP-2000-36\\
DESY 00-055\\
hep-ph/0004029
\end{flushright}
\vspace*{.5cm}
\vfill
\begin{center}
\boldmath
{\Large{\bf Effect of the Charm Quark Mass on the BFKL\\[.3cm]
$\gamma^* \gamma^* $ Total Cross Section at LEP$^*$}}
\unboldmath
\end{center}
\vspace{1cm}
\begin{center}
{\large
Jochen Bartels$^{a,1}$, Carlo Ewerz$^{b,c,2}$, Ren\'e Staritzbichler$^{a,3}$
}
\end{center}
\vspace{.2cm}
%\begin{center}
\begin{quote}
{\sl
$^a$ II.\ Institut f\"ur Theoretische Physik, Universit\"at Hamburg, 
Luruper Chaussee 149, D- 22761 Hamburg, Germany}\\[.2cm]
{\sl
$^b$ Cavendish Laboratory, Cambridge University, 
Madingley Road, Cambridge CB3 0HE, UK}\\[.2cm]
{\sl 
$^c$ DAMTP, Centre for Mathematical Sciences, Cambridge University, 
Wilberforce Road, Cambridge CB3 0WA, UK\\}
\end{quote}
%\end{center}
\vfill
\begin{abstract}
We perform a numerical study of various improvements of the LO BFKL 
prediction for the $\gamma^* \gamma^* $ total cross section in the 
kinematical region of the L3 detector at LEP. In particular, we study the 
effects of a massive charm quark and of different photon polarizations. 
The variation of the BFKL prediction under changes of $\alpha_s$ 
is investigated. 
\end{abstract}
\vfill
\vspace{5em}
\hrule width 5.cm
\vspace*{.5em}
{\small \noindent 
$^*$Work supported in part by the EU Fourth Framework Programme
`Training and Mobility of Researchers', Network `Quantum Chromodynamics
and the Deep Structure of Elementary Particles',
contract FMRX-CT98-0194 (DG 12 - MIHT).\\
$^1$email: {\sl bartels@x4u2.desy.de} \\
$^2$email: {\sl carlo@hep.phy.cam.ac.uk}\\ 
$^3$email: {\sl staritz@mail.desy.de} \\
}
\end{titlepage}

%1. 
\section{Introduction}

The total hadronic cross section in 
$\gamma^* \gamma^*$ scattering at electron--positron colliders 
is considered to be a suitable observable for studying the interesting 
dynamics of QCD at small $x$. For sufficiently large photon 
virtualities one expects this process to be the optimal test of 
the perturbative (BFKL) Pomeron \cite{BFKL}. 
The process is being studied by the L3 
and the OPAL collaborations \cite{l3,opal} at LEP. 
First analytic calculations based upon the leading order (LO) BFKL 
approximation \cite{BRL,BHS} have been compared to measurements 
by the L3 collaboration \cite{l3}. 
In the meantime various aspects of this process have been considered 
in more detail \cite{Wallon}--\cite{NLOint}. 
In summary, the present situation 
is similar to that of forward jets at HERA \cite{forwardex}. 
The data points for the total $\gamma^* \gamma^*$ cross section 
lie above the two--gluon exchange approximation 
but clearly below the LO BFKL prediction.
First attempts to include the NLO corrections \cite{FL,CC} (see also 
\cite{Gavinnlo}) to the BFKL approximation are encouraging 
\cite{Kwiecinski,NLOint} but not yet conclusive: 
for a consistent NLO calculation of the $\gamma^* \gamma^*$ cross section 
one also needs the NLO corrections to the photon impact factors which have 
not yet been calculated. 

But even the LO calculations still suffer from several theoretical 
uncertainties which we would like to discuss in this short note. 
The LO calculation of the  $\gamma^* \gamma^*$ 
cross section which has been compared with L3 data uses 
four massless quarks. Due to the charge $+2/3$ of the 
charm quark the cross section is multiplied by a factor 
of $2.8$ when going from three to four massless flavors. 
It is obviously very desirable to eliminate this big uncertainty. 
Effects of the charm mass have so far 
only been considered in \cite{Kwiecinski} where a formula 
was given in $x$-space and its effect was not considered 
separately. 
We give a formula for the cross section for non--zero charm 
quark mass in Mellin space and perform a numerical study 
which shows that the charm quark mass leads to a considerable 
reduction of the cross section expected for LEP.  
The theoretical prediction of the cross section also depends 
on the value of the strong coupling constant $\alpha_s$ 
which is not too well 
known at the (comparatively small) momentum scales 
dominating the kinematics at LEP, especially at 
$\sqrt{s}=91 \, \mbox{GeV}$. Our numerical study 
shows that the dependence on $\alpha_s$ is rather strong. 
Further, we briefly discuss the effect of taking into account 
contributions from all photon polarizations as well as 
the uncertainty associated with the BFKL energy scale. 

%We conclude that
%the LO BFKL prediction for LEP91 used in the L3 paper should be modified.

%2. 
\section{Cross section formula}
\label{formulasect}
In the events of interest the scattered electron as well as the 
scattered positron are tagged (`double--tag events'), and we 
define useful scaling variables (we use the notation of \cite{BRL}) 
\begin{equation}
  \label{x}
  x_1 = \frac{Q_1^2}{2q_1k_2}\,, \quad
  x_2 =  \frac{Q_2^2}{2q_2k_1}
\end{equation}
and 
\begin{equation}
  \label{y}
  y_1 =  \frac{q_1k_2}{k_1k_2}\,,\quad
  y_2 =  \frac{q_2k_1}{k_1k_2}\,, 
\end{equation}
where $k_1$ and $k_2$ are the momenta of the electron 
and positron, respectively. We have
\be
y_i = 1 -\frac{E^i_{tag}}{E_b}  \cos^2 
\left(\frac{\theta^i_{tag}}{2}\right)\,,
\ee
where $E_b$ is the beam energy, and $E^i_{tag}$ and $\theta^i_{tag}$ 
are the energy of the tagged lepton and its angle with respect to the beam 
axis, respectively. The virtualities of the photons are ($i=1,2$) 
\begin{equation}
\label{q}
Q_i^2  =  -q_i^2  =  2 E_b E^i_{tag} 
 (1-\cos \theta^i_{tag}) \,,
\end{equation}
and they are required to be large to make perturbation theory applicable. 
The squared center--of--mass energy of the $e^+e^-$ collisions is
$s=(k_1+k_2)^2$, 
whereas for the underlying $\gamma^*\gamma^*$ process the 
squared energy is given by
\begin{equation}
  \label{shat}
  \hat{s}  =  (q_1 + q_2)^2  \simeq  s y_1 y_2 \,. 
\end{equation}
We consider the limit where $Q_1^2,Q_2^2$ and $\hat{s}$ are large and
\begin{equation}
\label{lim}
  Q_1^2,  Q_2^2  \ll  \hat{s}\,.
\end{equation}
The differential $e^+e^-$ cross section has been calculated 
in \cite{BRL} and \cite{BHS}. 
The total $\gamma^*\gamma^*$ cross section is given by 
\begin{equation}
  \label{sum}
  \sigma_{\gamma^*\gamma^*} = \sigma_{\gamma^*\gamma^*}^{TT}
+\epsilon_2 \sigma_{\gamma^*\gamma^*}^{TL}
+\epsilon_1\sigma_{\gamma^*\gamma^*}^{LT}
+\epsilon_1\epsilon_2\sigma_{\gamma^*\gamma^*}^{LL}
\end{equation}
with
\begin{equation}
\quad \epsilon_i = \frac{ 2 (1-y_i)}{1 + (1-y_i)^2} \,.
\end{equation}
The labels $T$ and $L$ refer to the transverse and longitudinal 
polarization of the incoming photon, respectively. 
In \cite{l3} it has been argued that in the L3 region the $\epsilon_i$ 
have values larger than $0.97$. For simplicity we take them as 1 
which leads 
to a deviation of less than 3 percent. For the polarizations 
$i,j \in \{ L,T \}$ of the initial photons one finds
\be
  \label{gg}
  \sigma_{\gamma^*\gamma^*}^{ij}
= \frac {\alpha_{em}} {16\, Q_1^2}
  \frac {\alpha_{em}} {16 \,Q_2^2}
\int \frac{\d \nu}{2\pi^2} 
\exp \left[\hspace{1pt} \ln\left( \frac{\hat{s}}{s_0} \right) 
\chi (\nu) \right]  W_i \left( \nu, \frac{m^2}{Q_1^2} \right)
W_j \left(-\nu, \frac{m^2}{Q_2^2} \right) \,,
\ee
where $s_0$ is the typical BFKL energy scale which we choose as 
\be
\label{s0}
s_0 = \sqrt{Q_1^2Q_2^2}\,. 
\ee
Further, 
\be
\label{chi}
\chi(\nu) = N_c\alpha_s/\pi [ 2\psi(1) - \psi(1/2+i\nu) - \psi(1/2-i\nu)] \,,
\ee
and $\psi$ denotes the digamma function.
For massless quarks the functions $W_i$ 
are given in \cite{BRL}. 
For massive quarks we can use the results found in \cite{wuest} 
to obtain 
\begin{eqnarray}
\label{wqs}
 W_L \left(\nu,\frac{m_c^2}{Q_i^2}\right) & = &
               \sum\limits_{f=u,d,s}  8 \sqrt{2} \,q_f^2 \alpha_s\pi^2 
                 \frac{\nu^2+\frac{1}{4}}{\nu^2+1}\,
          \frac{\sinh \pi\nu}{\nu\cosh^2 \pi\nu} (Q_i^2)^{\frac{1}{2}+i\nu} \nn \\
       && +\,\frac{128}{15} \sqrt{2}\, q_c^2 \alpha_s\pi\sqrt{\pi} 
          \frac{\Gamma( \frac{3}{2} +i\nu )}{\Gamma (2+i\nu)\cosh \pi\nu }
          (Q_i^2)^{\frac{1}{2}+i\nu} 
          \left( \frac{Q_i^2 +4m_c^2}{Q_i^2} \right)^{-\frac{3}{2}-i\nu} \times 
   \nn \\
     &&\hspace{.3cm} 
\times \,  {}_2F_1 \left( \frac{3}{2} +i\nu, \frac{1}{2}; \frac{7}{2};
            \frac{Q_i^2}{Q_i^2 +4m_c^2} \right)
\end{eqnarray}
and 
\begin{eqnarray}
\label{w2}
\lefteqn{W_T \left(\nu,\frac{m_c^2}{Q_i^2}\right) = %} \nn \\
%&&   
\sum\limits_{f=u,d,s} 4 \sqrt{2} \,q_f^2 \alpha_s \pi^2 
                 \frac{\nu^2+\frac{9}{4}}{\nu^2+1}\,
          \frac{\sinh \pi\nu}{\nu\cosh^2 \pi\nu}(Q_i^2)^{\frac{1}{2}+i\nu} 
}
\nn\\   
&&   +\,16 \sqrt{2} \,q_c^2 \alpha_s\pi\sqrt{\pi} 
          \frac{\Gamma( \frac{1}{2} +i\nu )}{\Gamma (2+i\nu)\cosh \pi\nu }
          (Q_i^2)^{\frac{1}{2}+i\nu}
           \left( \frac{Q_i^2 +4m_c^2}{Q_i^2} \right)^{\frac{1}{2}-i\nu} 
 \times \nn \\
&&   \hspace{.3cm}
   \times \Biggr\{  
             \left[ 
               \frac{(1+3i\nu)Q_i^2+(3+2i\nu)m_c^2}{Q_i^2 +4m_c^2} 
               -i\nu\left( \frac{Q_i^2}{Q_i^2 +4m_c^2} \right)^2 
             \right]
                {}_2F_1 \left( \frac{1}{2} +i\nu, \frac{1}{2}; \frac{3}{2};
                  \frac{Q_i^2}{Q_i^2 +4m_c^2} \right) \nn \\ 
    &&  \hspace{1.2cm}
     +  \left[ (1-i\nu) \frac{Q_i^2}{Q_i^2 +4m_c^2}-
                  \frac{3+2 i\nu}{4} \right]
            {}_2F_1 \left(- \frac{1}{2} +i\nu, \frac{1}{2}; \frac{3}{2};
            \frac{Q_i^2}{Q_i^2 +4m_c^2} \right) 
        \Biggl\}
\end{eqnarray}
where $q_f$ denotes the electric charge of the quark flavor $f$. 
The terms containing $q_c =2/3$ correspond to the contribution 
of the charm quark. In the limit $m_c\to 0$ they can be simplified 
in such a way that the sum over flavours in the first terms is extended 
to include four flavours. 
%Here we have included the charm quark with $q_c=2/3$. 
The bottom quark --- relevant to experiments 
at a future Linear Collider --- can be included in the same way as 
the charm quark. (The contribution of the bottom quark is very 
small at LEP and will be neglected in the following.) 

\section{Numerical results}

For our numerical analysis we use the kinematics of the L3 detector 
(see also \cite{l3}). At LEP91 the outgoing electrons close to the 
forward direction can be detected within the angles
$30\,\mbox{mrad} < \theta_{tag} < 66\,\mbox{mrad}$, 
and the electron energy required for tagging is 
$E_{tag} > 30\,\mbox{GeV}$. 
This leads to a range of possible photon virtualities 
\be 
1.2 \,\mbox{GeV}^2 <  Q^2_{1,2}  <  9 \,\mbox{GeV}^2\,,
\ee
and the cross section is evaluated at the mean value 
$\langle Q^2 \rangle = 3.5 \,\mbox{GeV}^2$. 

At LEP183 the energy of the tagged electrons is restricted by
$E_{tag} > 40\, \mbox{GeV}$, 
and the angular range is the same as at LEP91.
The virtualities are
\be 
2.5 \,\mbox{GeV}^2 <  Q^2_{1,2}  <  35 \,\mbox{GeV}^2,
\ee
with the mean value
$\langle Q^2 \rangle = 14 \,\mbox{GeV}^2$.
Note that with these mean values of $Q^2$ the available range for the 
rapidity $Y= \ln (\hat{s}/s_0)$ is the same for LEP91 and LEP183. 

In the BFKL formalism the strong coupling $\alpha_s$ is kept 
fixed, and the natural scale is $\langle Q^2  \rangle$. 
In our numerical study, however, we use a running coupling 
in order to arrive at an improved prediction for the cross section. 
Specifically, we use a running $\alpha_s$ in NNLO with four 
active flavours. The variation of $\alpha_s$ at a given scale 
$\langle Q^2  \rangle$ (see below) is then obtained by 
varying $\Lambda_{\mbox{\tiny QCD}}$ (approximately between 
$100\,\mbox{MeV}$ and $350\,\mbox{MeV}$).

%3. 
The results of our numerical study are illustrated in the following figures. 
Fig.\ \ref{fig1a} shows a comparison between the saddle point 
approximation to the BFKL cross section 
and the exact calculation of the LO BFKL Pomeron, here 
for LEP183 and for transversely polarized photons only. 
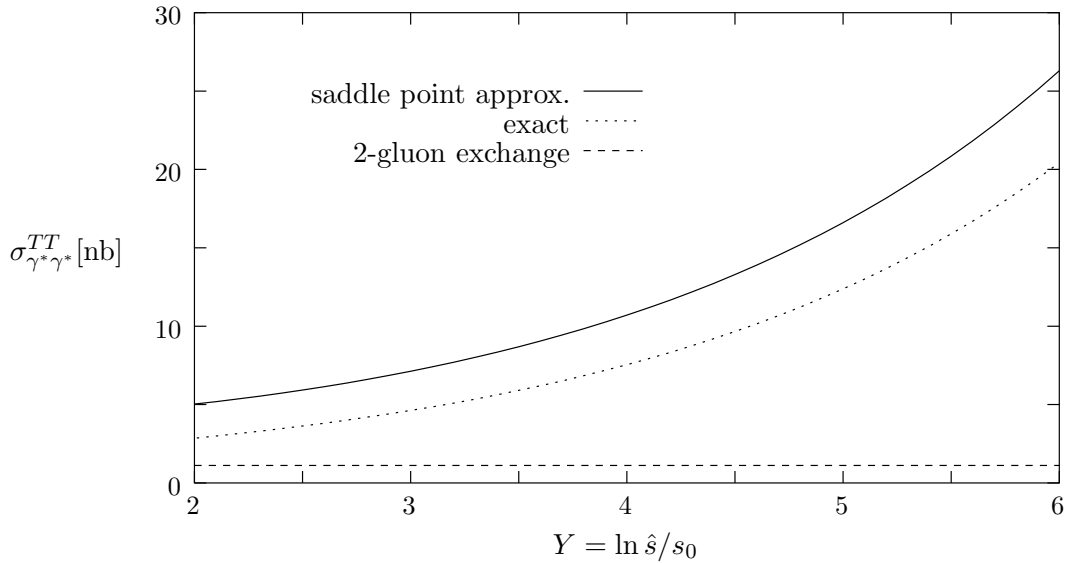
\begin{figure}
\begin{center}
\input{fig1a.pstex_t}
\caption{Comparison of the exact BFKL result 
with the saddle point approximation and the two--gluon 
exchange approximation for transversely 
polarized photons at LEP183 with  
$\langle Q^2 \rangle = 14\, \mbox{GeV}^2$, 
$\alpha_s= 0.208$}
\label{fig1a}
\end{center}
\end{figure}
In order to allow for a direct comparison with 
the theoretical curve used in \cite{l3}, the curves in 
this and in the next figure are calculated for four 
massless quarks (mass effects are discussed separately below). 
In \cite{l3} the measured data have been compared 
to the saddle point approximation. 
We find that this approximation 
overestimates the exact cross section by 20\% to 30\%. 
Fig.\ \ref{fig1a} also shows the cross section obtained 
from two--gluon exchange which 
approximates the full DGLAP formalism if the two 
photon virtualities are of the same order. 

\begin{figure}
\begin{center}
\input{fig1b.pstex_t}
\caption{
Contributions of different photon polarizations to the 
cross section, here 
for LEP 183 with $\langle Q^2 \rangle = 14\, \mbox{GeV}^2$, 
$\alpha_s = 0.208$, $m_c =0$}
\label{fig1b}
\end{center}
\end{figure}
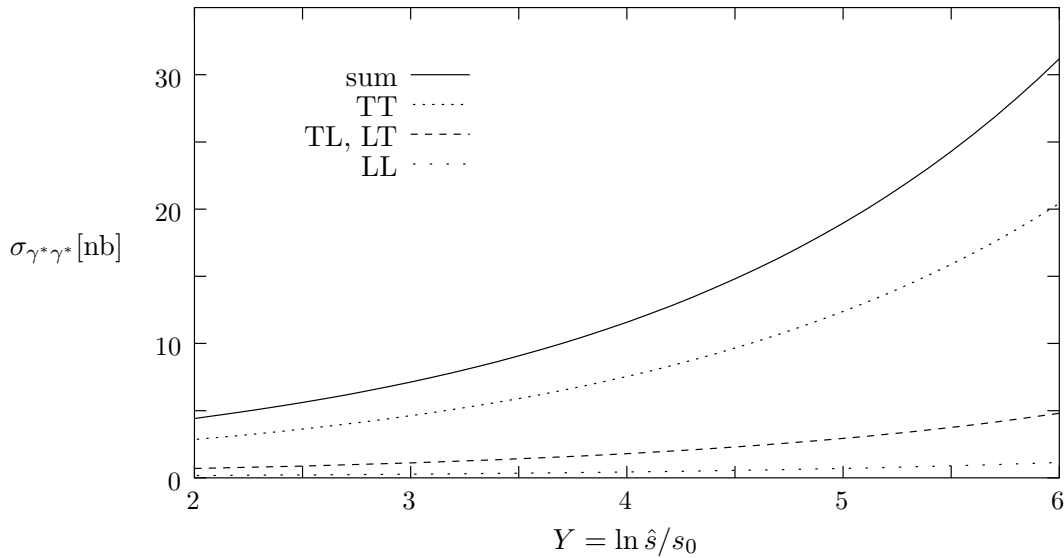

Next we show in Fig.\ \ref{fig1b} the contributions of different 
polarizations of the two incoming photons, again for LEP183 and 
with four massless quarks. Clearly the sum of all polarization is
substantially larger than the transverse contribution alone 
which was compared with data in \cite{l3}. 

Now we address the question how the cross section depends 
on the value of $\alpha_s$ chosen at the scale $\langle Q^2 \rangle$. 
Here and in the following we include the charm quark as a massive 
quark and sum over all photon polarizations, i.\,e.\ we use eq.\ 
(\ref{sum}) with (\ref{gg}), (\ref{wqs}), (\ref{w2}). 
As can be seen from eqs.\ (\ref{gg}) and (\ref{chi}), $\alpha_s$ 
enters in the exponent and thus we expect a strong dependence 
of the cross section on its particular value. 
This is confirmed by the numerical results presented in 
Figs.\ \ref{fig2a} and \ref{fig2b}. 
\begin{figure}
\begin{center}
\input{fig2a.pstex_t}
\caption{Energy dependence of the $\gamma^*\gamma^*$ cross section 
for different values of $\alpha_s(\langle Q^2 \rangle)$, here for 
LEP91 with $\langle Q^2 \rangle = 3.5 \,\mbox{GeV}^2$, 
$m_c = 1.5\, \mbox{GeV}$.}
\label{fig2a}
\end{center}
\end{figure}
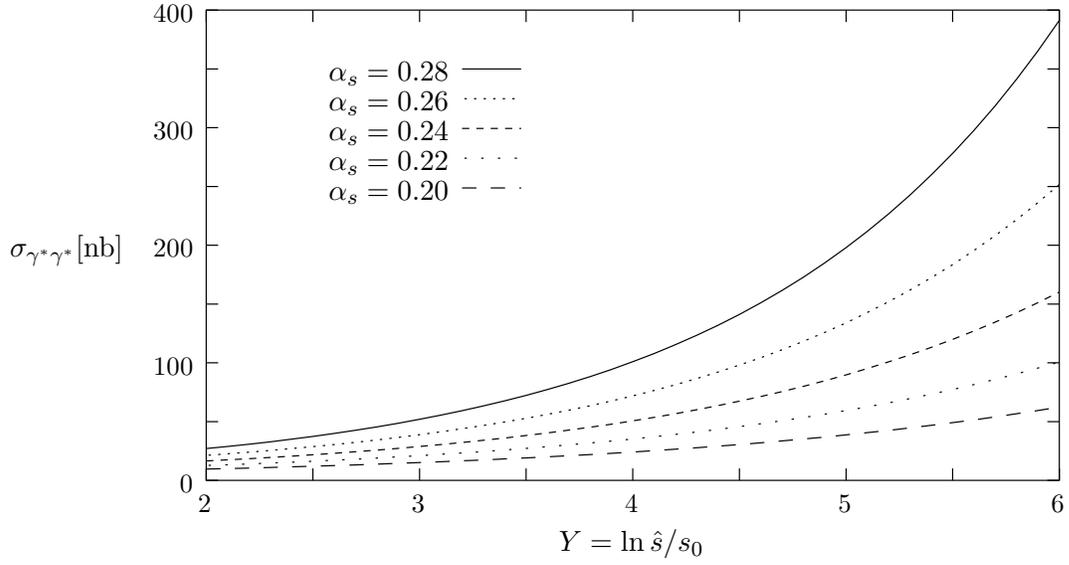
\begin{figure}
\begin{center}
\input{fig2b.pstex_t}
\caption{Same as Fig.\ \protect\ref{fig2a}, here for LEP183 with 
$\langle Q^2 \rangle = 14 \,\mbox{GeV}^2$}
\label{fig2b}
\end{center}
\end{figure}
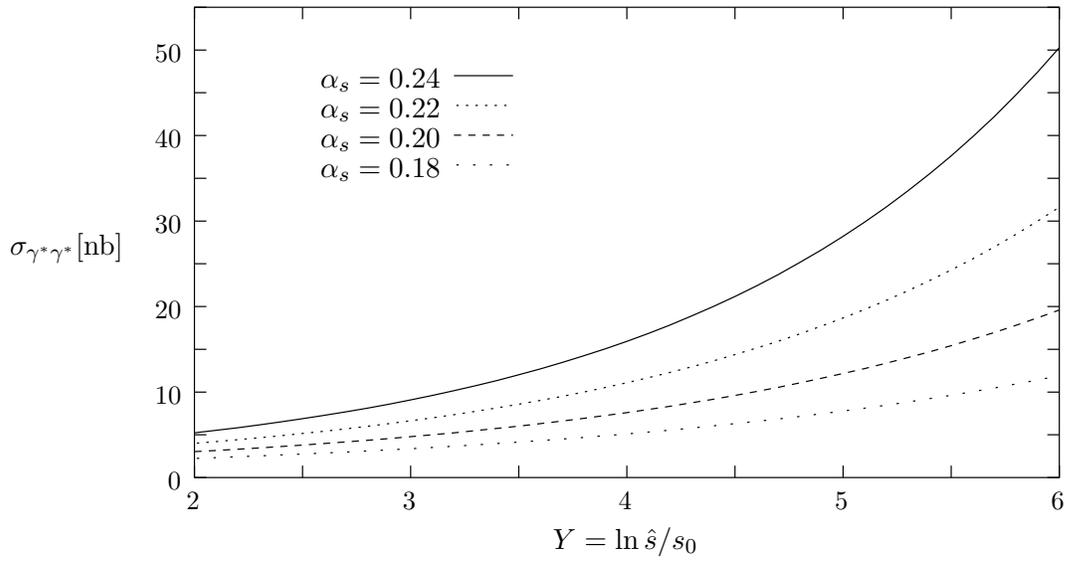
%keeping the number of flavors 
%fixed (4 flavors) we vary $\Lambda$  (approximately between $100$ MeV 
%and $350$ MeV).
Fig.\ \ref{fig2a} shows the cross section for 
$\langle Q^2 \rangle =3.5\,\mbox{GeV}^2$ (LEP91), 
and we study a spread $0.20<\alpha_s< 0.28$ of possible 
values. In Fig.\ \ref{fig2a} we have plotted the cross 
section for $\langle Q^2  \rangle =14\, \mbox{GeV}^2$ 
(LEP183), now varying $\alpha_s$ between $0.18$ and 
$0.24$. In \cite{l3} the theoretical curves were obtained 
using $\alpha_s=0.20$ for both energies. For LEP91 
this choice appears very low. To obtain more realistic 
predictions (see below), we choose $\alpha_s=0.268$ (LEP91) and 
$\alpha_s=0.208$ (LEP183), respectively. 

A further uncertainty is the choice of the BFKL energy scale $s_0$. 
Strictly speaking, its value is not determined in the leading order BFKL 
formalism. In this sense, the choice (\ref{s0}) is only an educated guess. 
It is well possible that the true value differs from that choice by 
a factor of two, for example. 
Choosing $s_0$ as twice (half) the value given in 
(\ref{s0}) results in a shift of the BFKL prediction to the right (left) 
by $\ln 2\simeq 0.7$ units in rapidity. Shifting the curves in our 
figures by this value amounts to a considerable change in the 
LO BFKL prediction. However, the uncertainty in $s_0$ is less 
serious in the NLO BFKL calculation since there the typical 
curves are less steep and a horizontal shift results in a smaller 
absolute change of the cross section. 

Finally, we study the effect of the charm quark mass on the 
cross section. Our results (including also the improvements 
discussed above) are shown in Fig.\ \ref{fig3a} for LEP91, 
and in Fig.\ \ref{fig3b} for LEP183. 
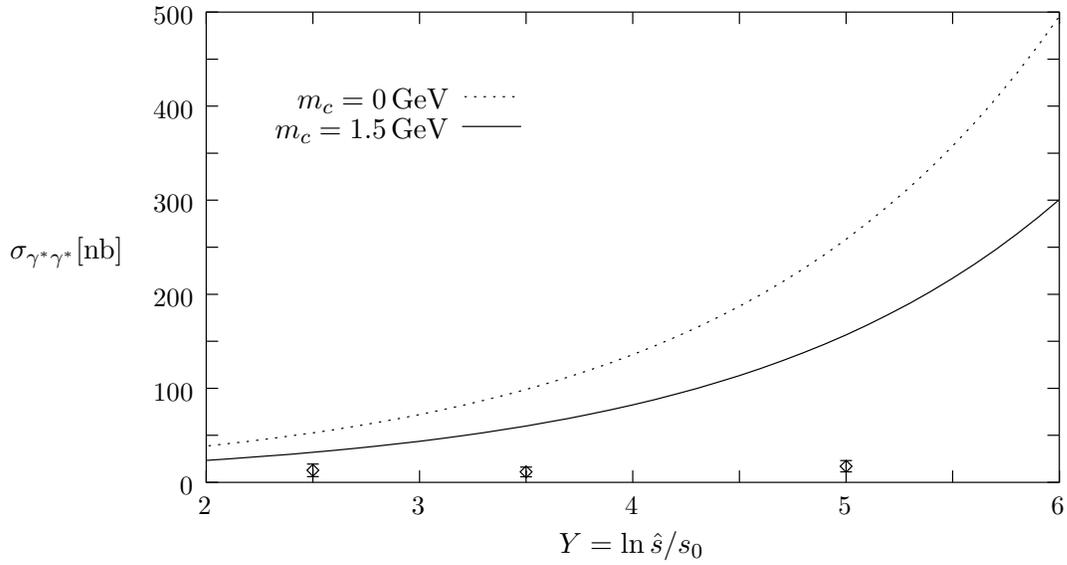
\begin{figure}
\begin{center}
\input{fig3a.pstex_t}
\caption{Effect of the charm quark mass on the 
$\gamma^*\gamma^*$ cross section for LEP 91 
with $\langle Q^2_i\rangle = 3.5 \,\mbox{GeV}^2, 
\alpha_s = 0.268$. Data points as measured by the 
L3 collaboration \protect\cite{l3}. }
\label{fig3a}
\end{center}
\end{figure}
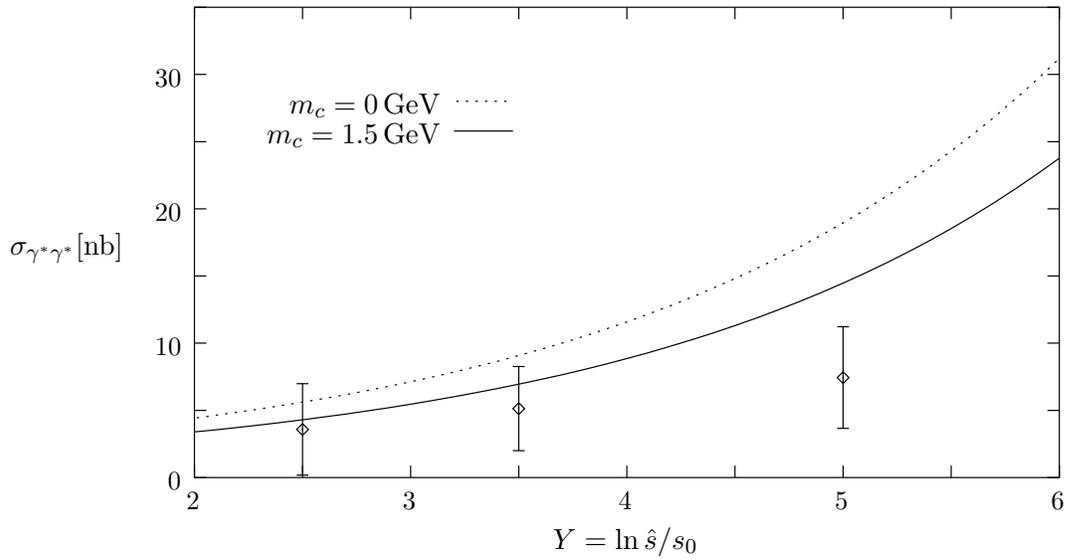
\begin{figure}
\begin{center}
\input{fig3b.pstex_t}
\caption{Same as Fig.\ \protect\ref{fig3a}, here for LEP183 with 
$\langle Q^2 \rangle = 14 \,\mbox{GeV}^2$, $\alpha_s = 0.208$}
\label{fig3b}
\end{center}
\end{figure}
For LEP91 the charm quark mass reduces the cross section by 
about 30 \% (compared to a massless charm quark), 
for LEP183 the cross section is reduced by about 20 \%. 
The effect of the charm mass is different in the two 
cases due to the different mean values $\langle Q^2 \rangle$. 
It can be seen immediately from eqs.\ (\ref{wqs}) and (\ref{w2}) 
that the charm mass $m_c$ enters in such a combination with 
$Q^2$ that for increasing $Q^2$ the effect of a finite $m_c$ 
becomes weaker. 
Also shown in Figs.\ \ref{fig3a} and \ref{fig3b} are the 
data points measured by the L3 collaboration \cite{l3}. 
One should note that in these data the contribution of the quark 
box diagram --- or quark parton model (QPM) graph ---
has already been subtracted. This diagram is also not 
included in our calculation. 

Compared with the theoretical prediction used in \cite{l3} 
we have made the following changes to arrive at the cross sections 
in Figs.\ \ref{fig3a} and \ref{fig3b}: 
exact BFKL is used instead of the saddle point approximation, 
all photon polarizations are included, and a nonzero 
charm quark mass is used. Further we have chosen a 
higher (and more realistic) value for $\alpha_s$ in the case 
of LEP91. Roughly speaking, the first three changes 
lead to a change in the normalization, whereas the last 
one results in a different energy dependence. 
For LEP91 we find a significantly larger cross section 
than the one given by the theoretical curve in \cite{l3}. Here 
the main effect is due to the choice of $\alpha_s$. 
For LEP183 our prediction is very close to the one 
used in \cite{l3} because the changes almost compensate 
each other. 
However, one should have in mind that already a small change 
in $\alpha_s$ leads to a visible modification of the energy 
dependence. 

Our predictions for the cross section are higher and exhibit a 
much faster rise with energy than the measured data points 
(especially in the case of LEP91). This was to be expected since 
the NLO corrections to the BFKL equation are known to 
be sizable and lead to a slower rise with energy. Certainly 
there is a good chance to successfully describe the data by 
a NLO BFKL calculation. However, also there one should 
keep in mind the uncertainties discussed above. 

%4. 
\section{Conclusions}
Our study of the BFKL prediction shows that there are 
substantial theoretical uncertainties even at the LO level. 
However, also an improved treatment does not change the 
main conclusion: the LEP data for the total hadronic 
$\gamma^*\gamma^*$ cross section clearly indicate that a 
simple two--gluon exchange is not sufficient to describe the 
data. The LO BFKL prediction, on the other hand, lies 
well above the data. A consistent NLO calculation 
including the impact factors is therefore urgently needed.  

\section*{Acknowledgements}
CE would like to thank G.\ Salam, M.\ Wadhwa and M.\ W\"usthoff 
for helpful discussions.

\end{document}

%% file: fig1a.pstex_t
\begin{picture}(0,0)%
\epsfig{file=fig1a.pstex}%
\end{picture}%
\setlength{\unitlength}{3947sp}%
\begingroup\makeatletter\ifx\SetFigFont\undefined%
\gdef\SetFigFont#1#2#3#4#5{%
  \reset@font\fontsize{#1}{#2pt}%
  \fontfamily{#3}\fontseries{#4}\fontshape{#5}%
  \selectfont}%
\fi\endgroup%
\begin{picture}(6398,3433)(543,-2999)
\put(1397,-2627){\makebox(0,0)[rb]{\smash{\SetFigFont{10}{12.0}{\familydefault}{\mddefault}{\updefault}0}}}
\put(1397,-1642){\makebox(0,0)[rb]{\smash{\SetFigFont{10}{12.0}{\familydefault}{\mddefault}{\updefault}10}}}
\put(1397,-656){\makebox(0,0)[rb]{\smash{\SetFigFont{10}{12.0}{\familydefault}{\mddefault}{\updefault}20}}}
\put(1397,329){\makebox(0,0)[rb]{\smash{\SetFigFont{10}{12.0}{\familydefault}{\mddefault}{\updefault}30}}}
\put(1471,-2751){\makebox(0,0)[b]{\smash{\SetFigFont{10}{12.0}{\familydefault}{\mddefault}{\updefault}2}}}
\put(2829,-2751){\makebox(0,0)[b]{\smash{\SetFigFont{10}{12.0}{\familydefault}{\mddefault}{\updefault}3}}}
\put(4187,-2751){\makebox(0,0)[b]{\smash{\SetFigFont{10}{12.0}{\familydefault}{\mddefault}{\updefault}4}}}
\put(5545,-2751){\makebox(0,0)[b]{\smash{\SetFigFont{10}{12.0}{\familydefault}{\mddefault}{\updefault}5}}}
\put(6903,-2751){\makebox(0,0)[b]{\smash{\SetFigFont{10}{12.0}{\familydefault}{\mddefault}{\updefault}6}}}
\put(4187,-2999){\makebox(0,0)[b]{\smash{\SetFigFont{11}{13.2}{\familydefault}{\mddefault}{\updefault}$Y = \ln \hat{s}/s_0$}}}
\put(3841,-164){\makebox(0,0)[rb]{\smash{\SetFigFont{11}{13.2}{\familydefault}{\mddefault}{\updefault}saddle point approx.}}}
\put(3841,-351){\makebox(0,0)[rb]{\smash{\SetFigFont{11}{13.2}{\familydefault}{\mddefault}{\updefault}exact}}}
\put(3841,-538){\makebox(0,0)[rb]{\smash{\SetFigFont{11}{13.2}{\familydefault}{\mddefault}{\updefault}2-gluon exchange}}}
\put(676,-1149){\makebox(0,0)[b]{\smash{\SetFigFont{11}{13.2}{\familydefault}{\mddefault}{\updefault}$\sigma^{TT}_{\gamma^*\gamma^*} [\mbox{nb}]$}}}
\end{picture}

%% file: fig1b.pstex_t
\begin{picture}(0,0)%
\epsfig{file=fig1b.pstex}%
\end{picture}%
\setlength{\unitlength}{3947sp}%
\begingroup\makeatletter\ifx\SetFigFont\undefined%
\gdef\SetFigFont#1#2#3#4#5{%
  \reset@font\fontsize{#1}{#2pt}%
  \fontfamily{#3}\fontseries{#4}\fontshape{#5}%
  \selectfont}%
\fi\endgroup%
\begin{picture}(6398,3402)(843,-2999)
\put(1697,-2627){\makebox(0,0)[rb]{\smash{\SetFigFont{10}{12.0}{\familydefault}{\mddefault}{\updefault}0}}}
\put(1697,-1782){\makebox(0,0)[rb]{\smash{\SetFigFont{10}{12.0}{\familydefault}{\mddefault}{\updefault}10}}}
\put(1697,-938){\makebox(0,0)[rb]{\smash{\SetFigFont{10}{12.0}{\familydefault}{\mddefault}{\updefault}20}}}
\put(1697,-93){\makebox(0,0)[rb]{\smash{\SetFigFont{10}{12.0}{\familydefault}{\mddefault}{\updefault}30}}}
\put(1771,-2751){\makebox(0,0)[b]{\smash{\SetFigFont{10}{12.0}{\familydefault}{\mddefault}{\updefault}2}}}
\put(3129,-2751){\makebox(0,0)[b]{\smash{\SetFigFont{10}{12.0}{\familydefault}{\mddefault}{\updefault}3}}}
\put(4487,-2751){\makebox(0,0)[b]{\smash{\SetFigFont{10}{12.0}{\familydefault}{\mddefault}{\updefault}4}}}
\put(5845,-2751){\makebox(0,0)[b]{\smash{\SetFigFont{10}{12.0}{\familydefault}{\mddefault}{\updefault}5}}}
\put(7203,-2751){\makebox(0,0)[b]{\smash{\SetFigFont{10}{12.0}{\familydefault}{\mddefault}{\updefault}6}}}
\put(4487,-2999){\makebox(0,0)[b]{\smash{\SetFigFont{11}{13.2}{\familydefault}{\mddefault}{\updefault}$Y = \ln \hat{s}/s_0$}}}
\put(3055,-93){\makebox(0,0)[rb]{\smash{\SetFigFont{11}{13.2}{\familydefault}{\mddefault}{\updefault}sum}}}
\put(3055,-280){\makebox(0,0)[rb]{\smash{\SetFigFont{11}{13.2}{\familydefault}{\mddefault}{\updefault}TT}}}
\put(3055,-467){\makebox(0,0)[rb]{\smash{\SetFigFont{11}{13.2}{\familydefault}{\mddefault}{\updefault}TL, LT}}}
\put(3055,-654){\makebox(0,0)[rb]{\smash{\SetFigFont{11}{13.2}{\familydefault}{\mddefault}{\updefault}LL}}}
\put(976,-1149){\makebox(0,0)[b]{\smash{\SetFigFont{11}{13.2}{\familydefault}{\mddefault}{\updefault}$\sigma_{\gamma^*\gamma^*} [\mbox{nb}]$}}}
\end{picture}

%% file: fig2a.pstex_t
\begin{picture}(0,0)%
\epsfig{file=fig2a.pstex}%
\end{picture}%
\setlength{\unitlength}{3947sp}%
\begingroup\makeatletter\ifx\SetFigFont\undefined%
\gdef\SetFigFont#1#2#3#4#5{%
  \reset@font\fontsize{#1}{#2pt}%
  \fontfamily{#3}\fontseries{#4}\fontshape{#5}%
  \selectfont}%
\fi\endgroup%
\begin{picture}(6398,3433)(843,-2999)
\put(1771,-2627){\makebox(0,0)[rb]{\smash{\SetFigFont{10}{12.0}{\familydefault}{\mddefault}{\updefault}0}}}
\put(1771,-1888){\makebox(0,0)[rb]{\smash{\SetFigFont{10}{12.0}{\familydefault}{\mddefault}{\updefault}100}}}
\put(1771,-1149){\makebox(0,0)[rb]{\smash{\SetFigFont{10}{12.0}{\familydefault}{\mddefault}{\updefault}200}}}
\put(1771,-410){\makebox(0,0)[rb]{\smash{\SetFigFont{10}{12.0}{\familydefault}{\mddefault}{\updefault}300}}}
\put(1771,329){\makebox(0,0)[rb]{\smash{\SetFigFont{10}{12.0}{\familydefault}{\mddefault}{\updefault}400}}}
\put(1845,-2751){\makebox(0,0)[b]{\smash{\SetFigFont{10}{12.0}{\familydefault}{\mddefault}{\updefault}2}}}
\put(3185,-2751){\makebox(0,0)[b]{\smash{\SetFigFont{10}{12.0}{\familydefault}{\mddefault}{\updefault}3}}}
\put(4524,-2751){\makebox(0,0)[b]{\smash{\SetFigFont{10}{12.0}{\familydefault}{\mddefault}{\updefault}4}}}
\put(5864,-2751){\makebox(0,0)[b]{\smash{\SetFigFont{10}{12.0}{\familydefault}{\mddefault}{\updefault}5}}}
\put(7203,-2751){\makebox(0,0)[b]{\smash{\SetFigFont{10}{12.0}{\familydefault}{\mddefault}{\updefault}6}}}
\put(4524,-2999){\makebox(0,0)[b]{\smash{\SetFigFont{11}{13.2}{\familydefault}{\mddefault}{\updefault}$Y = \ln \hat{s}/s_0$}}}
\put(3378,-41){\makebox(0,0)[rb]{\smash{\SetFigFont{11}{13.2}{\familydefault}{\mddefault}{\updefault}$\alpha_s = 0.28$}}}
\put(3378,-228){\makebox(0,0)[rb]{\smash{\SetFigFont{11}{13.2}{\familydefault}{\mddefault}{\updefault}$\alpha_s = 0.26$}}}
\put(3378,-415){\makebox(0,0)[rb]{\smash{\SetFigFont{11}{13.2}{\familydefault}{\mddefault}{\updefault}$\alpha_s = 0.24$}}}
\put(3378,-602){\makebox(0,0)[rb]{\smash{\SetFigFont{11}{13.2}{\familydefault}{\mddefault}{\updefault}$\alpha_s = 0.22$}}}
\put(3378,-789){\makebox(0,0)[rb]{\smash{\SetFigFont{11}{13.2}{\familydefault}{\mddefault}{\updefault}$\alpha_s = 0.20$}}}
\put(976,-1149){\makebox(0,0)[b]{\smash{\SetFigFont{11}{13.2}{\familydefault}{\mddefault}{\updefault}$\sigma_{\gamma^*\gamma^*} [\mbox{nb}]$}}}
\end{picture}

%% file: fig2b.pstex_t
\begin{picture}(0,0)%
\epsfig{file=fig2b.pstex}%
\end{picture}%
\setlength{\unitlength}{3947sp}%
\begingroup\makeatletter\ifx\SetFigFont\undefined%
\gdef\SetFigFont#1#2#3#4#5{%
  \reset@font\fontsize{#1}{#2pt}%
  \fontfamily{#3}\fontseries{#4}\fontshape{#5}%
  \selectfont}%
\fi\endgroup%
\begin{picture}(6398,3402)(1143,-2999)
\put(1997,-2627){\makebox(0,0)[rb]{\smash{\SetFigFont{10}{12.0}{\familydefault}{\mddefault}{\updefault}0}}}
\put(1997,-2090){\makebox(0,0)[rb]{\smash{\SetFigFont{10}{12.0}{\familydefault}{\mddefault}{\updefault}10}}}
\put(1997,-1552){\makebox(0,0)[rb]{\smash{\SetFigFont{10}{12.0}{\familydefault}{\mddefault}{\updefault}20}}}
\put(1997,-1015){\makebox(0,0)[rb]{\smash{\SetFigFont{10}{12.0}{\familydefault}{\mddefault}{\updefault}30}}}
\put(1997,-477){\makebox(0,0)[rb]{\smash{\SetFigFont{10}{12.0}{\familydefault}{\mddefault}{\updefault}40}}}
\put(1997, 60){\makebox(0,0)[rb]{\smash{\SetFigFont{10}{12.0}{\familydefault}{\mddefault}{\updefault}50}}}
\put(2071,-2751){\makebox(0,0)[b]{\smash{\SetFigFont{10}{12.0}{\familydefault}{\mddefault}{\updefault}2}}}
\put(3429,-2751){\makebox(0,0)[b]{\smash{\SetFigFont{10}{12.0}{\familydefault}{\mddefault}{\updefault}3}}}
\put(4787,-2751){\makebox(0,0)[b]{\smash{\SetFigFont{10}{12.0}{\familydefault}{\mddefault}{\updefault}4}}}
\put(6145,-2751){\makebox(0,0)[b]{\smash{\SetFigFont{10}{12.0}{\familydefault}{\mddefault}{\updefault}5}}}
\put(7503,-2751){\makebox(0,0)[b]{\smash{\SetFigFont{10}{12.0}{\familydefault}{\mddefault}{\updefault}6}}}
\put(4787,-2999){\makebox(0,0)[b]{\smash{\SetFigFont{11}{13.2}{\familydefault}{\mddefault}{\updefault}$Y = \ln \hat{s}/s_0$}}}
\put(3627,-101){\makebox(0,0)[rb]{\smash{\SetFigFont{11}{13.2}{\familydefault}{\mddefault}{\updefault}$\alpha_s = 0.24$}}}
\put(3627,-288){\makebox(0,0)[rb]{\smash{\SetFigFont{11}{13.2}{\familydefault}{\mddefault}{\updefault}$\alpha_s = 0.22$}}}
\put(3627,-475){\makebox(0,0)[rb]{\smash{\SetFigFont{11}{13.2}{\familydefault}{\mddefault}{\updefault}$\alpha_s = 0.20$}}}
\put(3627,-662){\makebox(0,0)[rb]{\smash{\SetFigFont{11}{13.2}{\familydefault}{\mddefault}{\updefault}$\alpha_s = 0.18$}}}
\put(1276,-1149){\makebox(0,0)[b]{\smash{\SetFigFont{11}{13.2}{\familydefault}{\mddefault}{\updefault}$\sigma_{\gamma^*\gamma^*} [\mbox{nb}]$}}}
\end{picture}

%% file: fig3a.pstex_t
\begin{picture}(0,0)%
\epsfig{file=fig3a.pstex}%
\end{picture}%
\setlength{\unitlength}{3947sp}%
\begingroup\makeatletter\ifx\SetFigFont\undefined%
\gdef\SetFigFont#1#2#3#4#5{%
  \reset@font\fontsize{#1}{#2pt}%
  \fontfamily{#3}\fontseries{#4}\fontshape{#5}%
  \selectfont}%
\fi\endgroup%
\begin{picture}(6398,3433)(1143,-2999)
\put(2071,-2627){\makebox(0,0)[rb]{\smash{\SetFigFont{10}{12.0}{\familydefault}{\mddefault}{\updefault}0}}}
\put(2071,-2036){\makebox(0,0)[rb]{\smash{\SetFigFont{10}{12.0}{\familydefault}{\mddefault}{\updefault}100}}}
\put(2071,-1445){\makebox(0,0)[rb]{\smash{\SetFigFont{10}{12.0}{\familydefault}{\mddefault}{\updefault}200}}}
\put(2071,-853){\makebox(0,0)[rb]{\smash{\SetFigFont{10}{12.0}{\familydefault}{\mddefault}{\updefault}300}}}
\put(2071,-262){\makebox(0,0)[rb]{\smash{\SetFigFont{10}{12.0}{\familydefault}{\mddefault}{\updefault}400}}}
\put(2071,329){\makebox(0,0)[rb]{\smash{\SetFigFont{10}{12.0}{\familydefault}{\mddefault}{\updefault}500}}}
\put(2145,-2751){\makebox(0,0)[b]{\smash{\SetFigFont{10}{12.0}{\familydefault}{\mddefault}{\updefault}2}}}
\put(3485,-2751){\makebox(0,0)[b]{\smash{\SetFigFont{10}{12.0}{\familydefault}{\mddefault}{\updefault}3}}}
\put(4824,-2751){\makebox(0,0)[b]{\smash{\SetFigFont{10}{12.0}{\familydefault}{\mddefault}{\updefault}4}}}
\put(6164,-2751){\makebox(0,0)[b]{\smash{\SetFigFont{10}{12.0}{\familydefault}{\mddefault}{\updefault}5}}}
\put(7503,-2751){\makebox(0,0)[b]{\smash{\SetFigFont{10}{12.0}{\familydefault}{\mddefault}{\updefault}6}}}
\put(4824,-2999){\makebox(0,0)[b]{\smash{\SetFigFont{11}{13.2}{\familydefault}{\mddefault}{\updefault}$Y = \ln \hat{s}/s_0$}}}
\put(3678,-203){\makebox(0,0)[rb]{\smash{\SetFigFont{11}{13.2}{\familydefault}{\mddefault}{\updefault}$m_c = 0 \,\mbox{GeV}$}}}
\put(3678,-390){\makebox(0,0)[rb]{\smash{\SetFigFont{11}{13.2}{\familydefault}{\mddefault}{\updefault}$m_c =  1.5 \,\mbox{GeV}$}}}
\put(1276,-1149){\makebox(0,0)[b]{\smash{\SetFigFont{11}{13.2}{\familydefault}{\mddefault}{\updefault}$\sigma_{\gamma^*\gamma^*} [\mbox{nb}]$}}}
\end{picture}

%% file: fig3b.pstex_t
\begin{picture}(0,0)%
\epsfig{file=fig3b.pstex}%
\end{picture}%
\setlength{\unitlength}{3947sp}%
\begingroup\makeatletter\ifx\SetFigFont\undefined%
\gdef\SetFigFont#1#2#3#4#5{%
  \reset@font\fontsize{#1}{#2pt}%
  \fontfamily{#3}\fontseries{#4}\fontshape{#5}%
  \selectfont}%
\fi\endgroup%
\begin{picture}(6398,3402)(843,-2999)
\put(1697,-2627){\makebox(0,0)[rb]{\smash{\SetFigFont{10}{12.0}{\familydefault}{\mddefault}{\updefault}0}}}
\put(1697,-1782){\makebox(0,0)[rb]{\smash{\SetFigFont{10}{12.0}{\familydefault}{\mddefault}{\updefault}10}}}
\put(1697,-938){\makebox(0,0)[rb]{\smash{\SetFigFont{10}{12.0}{\familydefault}{\mddefault}{\updefault}20}}}
\put(1697,-93){\makebox(0,0)[rb]{\smash{\SetFigFont{10}{12.0}{\familydefault}{\mddefault}{\updefault}30}}}
\put(1771,-2751){\makebox(0,0)[b]{\smash{\SetFigFont{10}{12.0}{\familydefault}{\mddefault}{\updefault}2}}}
\put(3129,-2751){\makebox(0,0)[b]{\smash{\SetFigFont{10}{12.0}{\familydefault}{\mddefault}{\updefault}3}}}
\put(4487,-2751){\makebox(0,0)[b]{\smash{\SetFigFont{10}{12.0}{\familydefault}{\mddefault}{\updefault}4}}}
\put(5845,-2751){\makebox(0,0)[b]{\smash{\SetFigFont{10}{12.0}{\familydefault}{\mddefault}{\updefault}5}}}
\put(7203,-2751){\makebox(0,0)[b]{\smash{\SetFigFont{10}{12.0}{\familydefault}{\mddefault}{\updefault}6}}}
\put(4487,-2999){\makebox(0,0)[b]{\smash{\SetFigFont{11}{13.2}{\familydefault}{\mddefault}{\updefault}$Y = \ln \hat{s}/s_0$}}}
\put(3327,-262){\makebox(0,0)[rb]{\smash{\SetFigFont{11}{13.2}{\familydefault}{\mddefault}{\updefault}$m_c =0 \,\mbox{GeV}$}}}
\put(3327,-449){\makebox(0,0)[rb]{\smash{\SetFigFont{11}{13.2}{\familydefault}{\mddefault}{\updefault}$m_c = 1.5 \,\mbox{GeV}$}}}
\put(976,-1149){\makebox(0,0)[b]{\smash{\SetFigFont{11}{13.2}{\familydefault}{\mddefault}{\updefault}$\sigma_{\gamma^*\gamma^*} [\mbox{nb}]$}}}
\end{picture}

%% file: paper.bbl
\begin{thebibliography}{100}
\bibitem{BFKL}
E.\,A.\ Kuraev, L.\,N.\ Lipatov, V.\,S.\ Fadin,
%               Sov. {\bf JETP 44} (1976) 443; \\
{\it Sov.\ Phys.\ JETP\/ }{\bf 45} (1977) 199;\\
Ya.\,Ya.\ Balitskii, L.\,N.\ Lipatov, 
{\it Sov.\ J.\ Nucl.\ Phys.\ }{\bf 28} (1978) 822
\bibitem{l3}
L3 Collaboration, M.\ Acciarri et al., 
{\it Phys.\ Lett.\ }{\bf B 453} (1999) 333
\bibitem{opal}
M.\ Przybycie\'n for the OPAL collaboration, 
{\it Nucl.\ Phys.\ }{\bf B} {\it(Proc.\ Suppl.)} {\bf 82} (2000) 67
\bibitem{BRL}
J.\ Bartels, A.\ De Roeck, H.\ Lotter, 
{\it Phys.\ Lett.\ }{\bf B 389} (1996) 742 [hep-ph/9608401];\\
J.\ Bartels, A.\ De Roeck, C.\ Ewerz, H.\ Lotter, hep-ph/9710500, 
in `ECFA/DESY Study on Physics and Detectors for a Linear Collider', 
ed.\ R.\ Settles, DESY 97-123E 
\bibitem{BHS} 
S.\,J.\ Brodsky, F.\ Hautmann, D.\,E.\ Soper, 
{\it Phys.\ Rev.\ Lett.\ }{\bf 78} (1997) 803, 
erratum {\it ibid.\ }{\bf 79} (1997) 3522 
[hep-ph/9610260];\\
S.\,J.\ Brodsky, F.\ Hautmann, D.\,E.\ Soper, 
{\it Phys.\ Rev.\ }{\bf D 56} (1997) 6957 [hep-ph/9706427] 
\bibitem{Wallon}
M.\ Boonekamp, A.\ De Roeck, C.\ Royon, S.\ Wallon, 
{\it Nucl.\ Phys.\ }{\bf B 555} (1999) 540 [hep-ph/9812523]
\bibitem{bialas}
A.\ Bialas, W.\ Czyz, W.\ Florkowski, 
{\sl Eur.\ Phys.\ J.\ }{\bf C 2} (1998) 683, [hep-ph/9705470] 
\bibitem{Nikolaev}
N.\,N.\ Nikolaev, J.\ Speth, V.\,R.\ Zoller, 
hep-ph/0001120 
\bibitem{Levin}
E.\ Gotsman, E.\ Levin, U.\ Maor, E.\ Naftali, 
hep-ph/0001080
\bibitem{Dosch}
A.\ Donnachie, H.\,G.\ Dosch, M.\ Rueter, 
{\sl Eur.\ Phys.\ J.\ }{\bf C 13} (2000) 141 [hep-ph/9908413]
\bibitem{Kwiecinski}
J.\ Kwieci\'nski, L.\ Motyka, 
{\it Phys.\ Lett.\ }{\bf B 462} (1999) 203 [hep-ph/9905567]
\bibitem{DSR}
A.\ Donnachie, S.\ S{\"o}ldner-Rembold, hep-ph/0001035, 
to be published in the proceedings of UK Phenomenology 
Workshop on Collider Physics, Durham 1999
\bibitem{NLOint} 
S.\,J.\ Brodsky, V.\,S.\ Fadin, V.\,T.\ Kim, L.\,N.\ Lipatov, 
G.\,B.\ Pivovarov, 
{\it JETP Lett.\ }{\bf 70} (1999) 155 [hep-ph/9901229]
%Pisma ZhETF {\bf 70}, (1999) 161
\bibitem{forwardex}
H1 Collaboration, C.\ Adloff et al., 
{\it Nucl.\ Phys.\ }{\bf B 538} (1999) 3 [hep-ex/9809028];\\
ZEUS Collaboration, J.\ Breitweg et al.,
{\sl Eur.\ Phys.\ J.\ }{\bf C 6} (1999) 239 [hep-ex/9805016] 
\bibitem{FL} 
V.\,S.\ Fadin, L.\,N.\ Lipatov, 
{\it Phys.\ Lett.\ }{\bf B 429} (1998) 127 [hep-ph/9802290] 
and references therein 
\bibitem{CC}
M.\ Ciafaloni, G.\ Camici, 
{\it Phys.\ Lett.\ }{\bf B 430} (1998) 349 [hep-ph/9803389] 
and references therein
\bibitem{Gavinnlo}
M.\ Ciafaloni, D.\ Colferai and G.\,P.\ Salam, 
{\it Phys.\ Rev.\ }{\bf D 60} (1999) 114036 [hep-ph/9905566] 
\bibitem{wuest}
K.\ Golec-Biernat, M.\ W\"usthoff, 
{\it Phys.\ Rev.\ }{\bf D 59} (1999) 014017 [hep-ph/9807513]
\end{thebibliography}
